\documentclass[12pt]{iopart}
\usepackage{graphicx}
\usepackage{amsmath,amssymb}
\usepackage{dcolumn}
\usepackage{color}
\usepackage{hyperref} 
\usepackage{bm}
\usepackage{xcolor}
\usepackage[maxfloats=256]{morefloats}
\begin{document}

\title[MQM with Yb in the $^3P_2$ state ]{Measuring the nuclear magnetic quadrupole moment of optically trapped ytterbium atoms in the metastable state}
\author{Ayaki Sunaga$^{1*}$ \footnote{Present address: ELTE, E\"otv\"os Lor\'and University, Institute of Chemistry, P\'azm\'any P\'eter s\'et\'any 1/A 1117 Budapest, Hungary}, 
Yuiki Takahashi$^2$, Amar Vutha$^3$ and Yoshiro Takahashi$^1$}
\address{$^1$Department of Physics, Graduate School of Science, Kyoto University, Kyoto 606-8502, Japan \\
$^2$Division of Physics, Mathematics, and Astronomy, California Institute of Technology, Pasadena, California 91125, USA \\
$^3$Department of Physics, University of Toronto, Toronto M5S 1A7, Canada 
}
\ead{sunagaayaki@gmail.com}
\vspace{10pt}

\begin{abstract}
 We propose a scheme to measure a nuclear magnetic quadrupole moment (MQM), a $\mathcal{CP}$-violating electromagnetic moment that appears in the nuclear sector, using the long-lived $^3P_2$ metastable state in neutral $^{173}$Yb atoms. Laser-cooling and trapping techniques enable us to prepare ultracold $^{173}$Yb atoms in the $^3P_2$ state trapped in an optical lattice or an optical tweezer array, providing an ideal experimental platform with long spin coherence time. In addition, our relativistic configuration interaction calculation for the $^3P_2$ electronic wavefunction reveals a large magnetic field gradient generated by the atomic electrons in this state, which amplifies the measurable effect of an MQM.
Our scheme could lead to an improvement of more than one order of magnitude in MQM sensitivity compared to the best previous measurement [S. A. Murthy $et$ $al.$, Phys. Rev. Lett. \textbf{63}, 965 (1989)].
\end{abstract}

%
\vspace{2pc}

---------
\maketitle

%
%
\section{Introduction}\label{sec:introduction}
While the standard model (SM) of particle physics is a well-established model \cite{Gaillard1999RMP_SM}, the violation of the symmetries of charge conjugation and parity ($\mathcal{CP}$) that emerges in the Cabibbo–Kobayashi–Maskawa matrix in the SM is too small to explain the baryon asymmetry of the universe \cite{Huet1995PD_BAU,Dine2003RMP_BAU,Canetti2012NJP_BAU}, one of the unsolved questions in modern physics. This motivates the search for sources of larger $\mathcal{CP}$ violations that can be described by physics beyond the SM. Under the assumption of $\mathcal{CPT}$ invariance ($\mathcal{T}$ denoting time reversal), $\mathcal{CP}$ violation implies the violation of $\mathcal{T}$ symmetry.

The effect of $\mathcal{CP}$ violation in elementary particles can appear in low-energy-high-resolution measurements using atoms and molecules \cite{Pospelov2005AP,Safronova2018RMP,Chupp2019RMP}.
An electric dipole moment of an electron (eEDM) probes $\mathcal{CP}$-violation in the leptonic sector, whereas the nuclear Schiff moment and magnetic quadrupole moment (MQM) probe $\mathcal{CP}$-violation in the hadronic sector.
Measurements from these two sectors are complementary since they are sensitive to different types of new physics that cause $\mathcal{CP}$-violation \cite{Pospelov2005AP,Chupp2019RMP}. 
%
Despite the excellent precision and accuracy of Schiff moment measurements using Hg \cite{Hg2009_EDM,Hg2016_EDM} and Xe \cite{Xe2019_EDM_PRL,Xe2019_EDM_PRA}, there remains some ambiguity in relating the measured energy shifts to microscopic $\mathcal{CP}$ violation parameters. 
Nuclear many-body calculations are required to determine the upper limit of $\mathcal{CP}$-violating parameters, such as proton's EDM. However, the discrepancy between the calculations reported is considerable as even the signs of the values may not agree \cite{Engel2013PPNP_nuc_phys}, which affects the determination of the upper limit of $\mathcal{CP}$-violation parameters \cite{Hg2016_EDM,Xe2019_EDM_PRA}. Broadly, this difficulty arises because the Schiff moment is a collective nuclear property, which becomes measurable only because of the polarization of the electron cloud in an atom by the finite-sized nucleus \cite{Sushkov1984JETP_MQM,Flambaum1986NP}. 
%
%
In contrast, the contributions of $\mathcal{CP}$-violating sources in the MQM can be estimated more accurately because single nucleon $\mathcal{P}$- and $\mathcal{T}$-violating effects directly contributes to the MQM \cite{Flambaum2014PRL_MQM,Lackenby2018PRA_MQM}. 
We note that the MQM has been investigated not only at the nucleon level
, but also at the quark level \cite{Liu2012PLB}. 
Importantly, the MQM is enhanced in some heavily deformed nuclei, as has been confirmed using spherical basis calculations  \cite{Flambaum2014PRL_MQM,Flambaum1994PLB_MQM} and Nilsson model calculations \cite{Lackenby2018PRA_MQM}. 
While there are several nuclei that can be used for MQM measurements, including some radioactive nuclei, there has been considerable interest in the stable isotope $^{173}$Yb, which has $I=5/2$ and large quadrupole deformation that leads to huge enhancement of the MQM \cite{Moller2016ADNDT_deformation, Ho2023FP}. This fact has led to experimental proposals to measure the MQM using molecules like YbF \cite{Ho2023FP} and YbOH \cite{Takahashi2023arXiv,Pilgram2021JCP_YbOH173}, exploiting the internal magnetic field gradient induced by the electric field in polarized molecules. Recently, MQM measurements using solids containing deformed nuclei have also been proposed \cite{ICAP_Vutha_Eu,Dalton2023,Ramachandran2023a_Eu}.


In this work, we present a scheme to measure the MQM of $^{173}$Yb using ultracold $^{173}$Yb atoms in the long-lived metastable state $^3P_2$, and accurately compute the MQM sensitivity of this system. Our relativistic configuration interaction theory calculation for the $^3P_2$ electronic wavefunction reveals a large magnetic field gradient, a key ingredient for the magnetic quadrupole moment measurement. The preparation of ultracold Yb atoms in the $^3P_2$ state in an optical lattice \cite{Tomita2019PRA_2body_loss} and an optical tweezer array \cite{Okuno2022JPSP} has already been demonstrated, providing an ideal experimental platform with long spin coherence time for the MQM measurement.

\begin{figure*}[ht]
\includegraphics[width=150mm]{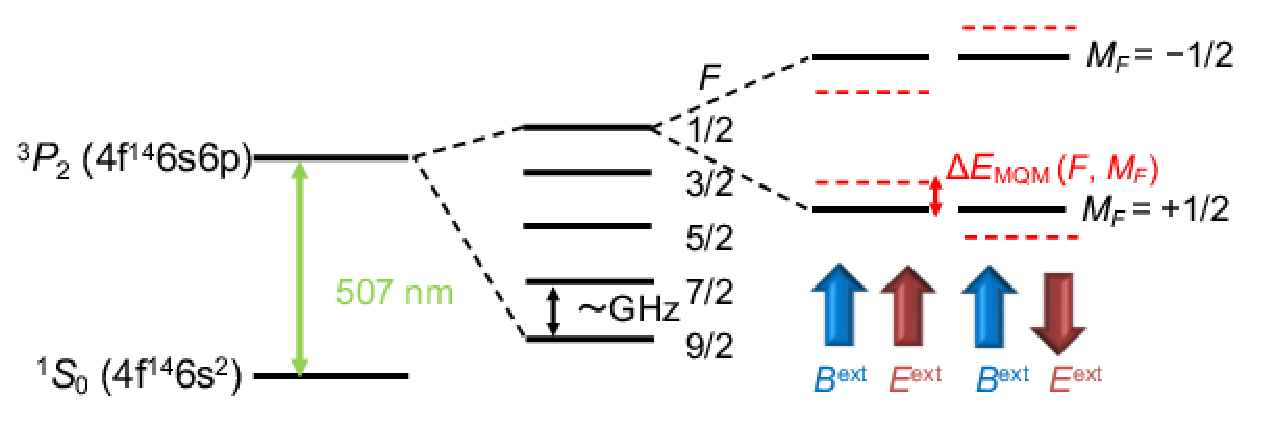}
\caption{\label{fig:overview} Overview of experiment to search for the MQM using an $^{173}$Yb ($I=5/2$) atom in the $^3P_2$ state. 
The energy shifts in the hyperfine levels are induced by the MQM as a result of the application of a DC electric field ($\mathcal{E}^\mathrm{ext}$) parallel and anti-parallel to the magnetic field ($\mathcal{B}^\mathrm{ext}$).
The difference in energy shifts between the two cases signals the MQM.
This figure only describes the cases of the $\left|F=1/2,M_F=+1/2\right>$ and $\left|F=1/2,M_F=-1/2\right>$ states, but others can also be employed for the MQM search.}
\end{figure*}

Figure \ref{fig:overview} shows the overview of the proposed experiment.
%
%
%
%
%
\begin{figure*}[ht]
\centering
\includegraphics[width=\columnwidth]{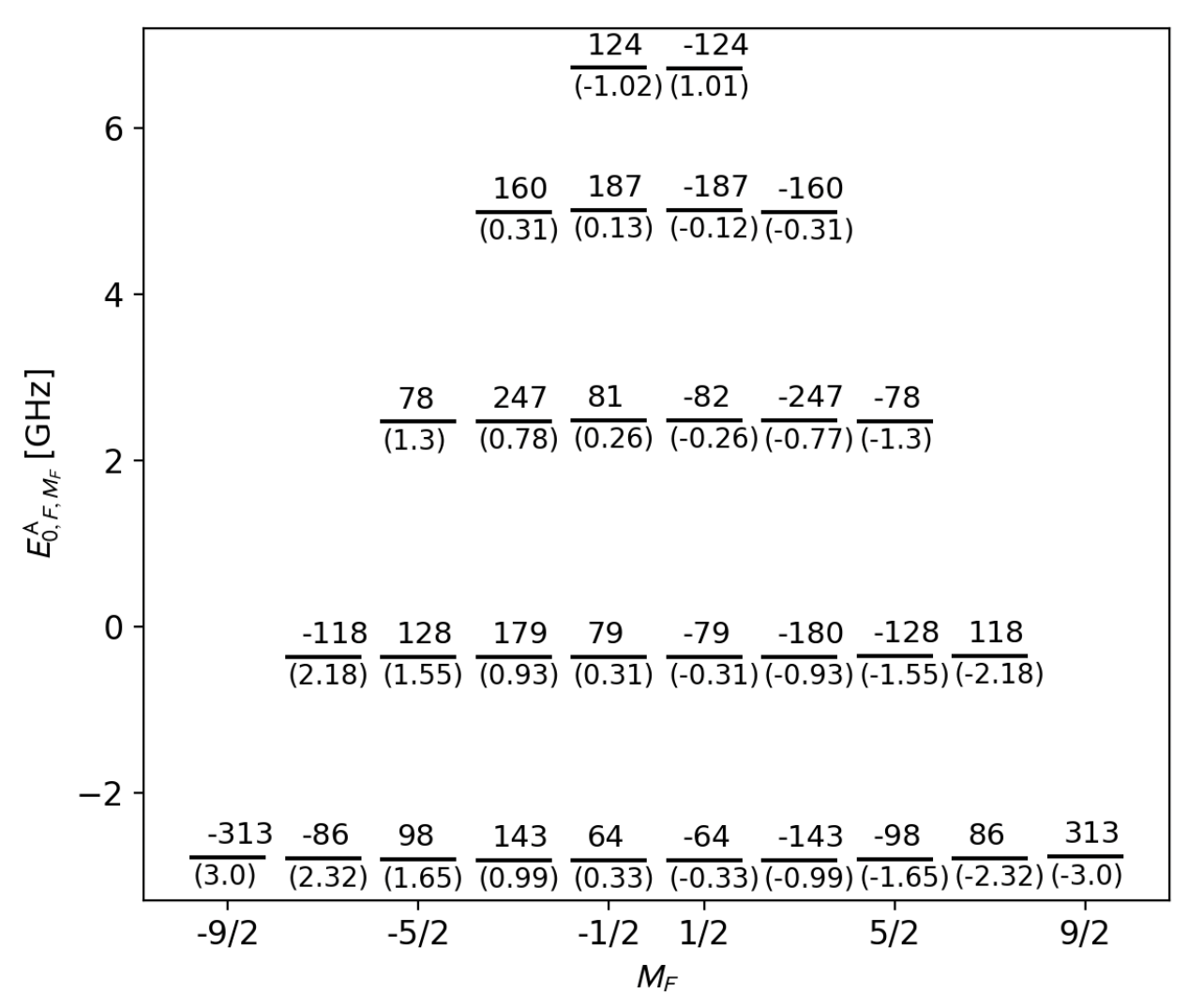}
\caption{\label{fig:MQM_all} Energy shifts of the hyperfine states in the $^3P_2$ state of $^{173}$Yb due to the MQM interaction, $\Delta E_\mathrm{MQM}({F,M_F})$, in units of Hz (above the states). These energy shifts are for an MQM magnitude $\mathcal{M}$ = 1 $\mu_N\mathrm{fm}$ and an applied electric field $\mathcal{E}_z$ = 50 kV/cm. As a measure of the sensitivity of a hyperfine state to magnetic field noise and systematic errors, the magnetic moments, $\mu_{F,M_F}$, in units of $\mu_B$ given by Eq. \ref{eq:g_F} (below the states in parentheses) are shown for an applied magnetic field $\mathcal{B}_z$ = 1 G.}
\end{figure*}
%
Figure \ref{fig:MQM_all} summarizes the main results of this work. Energy shifts, produced by an MQM of $\mathcal{M} = 1$ $\mu_N\mathrm{fm}$  under an electric field of 50 kV/cm and a magnetic field of 1 G, are shown for all hyperfine levels in the $^3P_2$ state. Since magnetic field drifts and noise can affect the measurement of the small shifts due to an MQM, we have also indicated the magnetic moment of each state in the figure.

From the information in Fig. \ref{fig:MQM_all}, we estimate the MQM sensitivity obtained from this scheme based on a typical setup of the optical lattice or optical tweezer array as $\delta \mathcal{M} = 5.2\times 10^{-8}$ $\mu_N\mathrm{fm}$, indicating sufficient precision to improve upon the current experimental bound obtained with $^{133}$Cs atoms \cite{Murthy1989PRL_Cs}. 
We also discuss a way to mitigate the deleterious effects of fluctuating magnetic fields during the experiment, by tuning the experimental parameters such that a chosen pair of hyperfine states has large differential MQM sensitivity but nearly identical magnetic moments. 


This paper is organized as follows: In Sec. \ref{sec:theory} we explain the relevant theory for the atomic energy level shifts due to the nuclear MQM. The computational details for the internal magnetic field gradient are described, along with the details of the calculation of MQM sensitivities for the hyperfine states \cite{zenodo:data_mqm_explain}. The proposed experimental scheme for the MQM measurement is presented in Sec. \ref{sec:experiment}. Section \ref{sec:conclusion} is devoted to conclusions. 
%
%
\section{Theory}\label{sec:theory}
\subsection{Nuclear magnetic quadrupole moment}\label{subsec:MQM}
Similar to the nuclear electric quadrupole interaction \cite{Ramsey1953_nucmoments}, the Hamiltonian for the nuclear magnetic quadrupole interaction is given by
%
\begin{equation}\label{eq:MQM}
\hat{H}_{\mathrm{MQM}} =-\sum_{i,j} \frac{1}{6} \mathcal{M}_{i j}(\nabla \mathbf{\mathcal{B}})_{i j},
\end{equation}
where $i$ and $j$ are Cartesian coordinate indices, $\mathcal{M}_{ij}$ is the $i,j$ component of the magnetic quadrupole moment tensor, and $\mathbf{\mathcal{B}}$ is the magnetic field. As the magnetic quadrupole moment is extremely small, it is sufficient to consider just the diagonal matrix elements of this Hamiltonian which lead to first-order perturbations. Thus the effective Hamiltonian for a nuclear spin, $I$, quantized along the $z$-axis can be expressed (with $i=j=z$) as
%
\begin{eqnarray}\label{eq:MQM_MI}
\hat{H}_{\mathrm{MQM}} & \rightarrow & -\frac{1}{6} \mathcal{M}_{z z}(\nabla \mathbf{B})_{z z} \\
& = & -\frac{1}{6} \frac{\mathcal{M}}{I(2 I-1)}\left[3 {M_I}^2-I(I+1)\right](\nabla \mathbf{B})_{zz}. \nonumber
\end{eqnarray}
Here $\mathcal{M}$ is the magnitude of the MQM, $M_I$ is the projection of $I$ along the quantization axis $z$. Inside an atom, the gradient of the magnetic field $(\nabla \mathbf{B})_{i j}$ is produced by the spins and orbital motion of the atomic electrons. In the four-component relativistic framework, the magnetic field gradient in atomic units is
%
\begin{equation}\label{eq:MFG}
(\nabla \mathbf{B})_{zz} = 
\sum_i \frac{\partial}{\partial z}\left(\frac{1}{c}\frac{\boldsymbol{\alpha}_i\times \mathbf{r}_i  }{r_i^3}\right)_z =
\frac{3}{c}\sum_i \left(\frac{\mathbf{r}_i \times \boldsymbol{\alpha}_i }{r_i^5}\right)_zz_i, 
\end{equation}
$ \bm{\alpha}$ is the Dirac matrix,
and $r$ is the distance of an electron from the center of the nucleus. 
%
%
\subsection{Atomic energy level shifts induced by the nuclear MQM}\label{subsec:formulation}
While the nuclear MQM itself has no direct interaction with an electric field, MQM measurements are carried out by measuring energy shifts in the presence of a static electric field, as shown in Fig. \ref{fig:overview} and \ref{fig:MQM_all}. This is because the magnetic field gradient is a $\mathcal{P}$-odd quantity, and a nonzero magnetic field gradient in an atom is only generated after parity mixing between the electronic orbitals in an externally-applied electric field. 
In this section, we formally show how the energy shifts due to the MQM arise in the electric field. The discussion is along similar lines to the case of eEDM-induced energy shifts \cite{Sandars1968JPB_formulation,Nataraj_thesis,Mukherjee2009JPCA}.
%
%
\subsubsection{The case with no hyperfine coupling}\label{subsubsec:non_IJ}
First, for simplicity, we ignore the hyperfine coupling between $\textbf{I}$ and $\textbf{J}$, and thus the product states $\left|JM_J\right>\left|IM_I\right>$ are the eigenstates. We express the contribution from $\hat{H}_{\mathrm{MQM}}$ in the perturbation theory. 
The eigenvalue problem of the total Hamiltonian ($\hat{H}^{\mathrm{e}}$) can be written as
\begin{equation}
\hat{H}^\mathrm{e}\left|\Psi^\mathrm{e}_m\right\rangle = E^\mathrm{e}_m\left|\Psi^\mathrm{e}_m\right\rangle, 
\end{equation}
where $E^\mathrm{e}_m$ and $\left|\Psi^\mathrm{e}_m\right\rangle$ are the energy eigenvalue and eigenstate, respectively. We focus on the $^3P_2$ state for $\left|\Psi^{\mathrm{e}}_m\right>$ state in this study.
The total Hamiltonian $\hat{H}^{\mathrm{e}}$ can be separated into the unperturbed ($\hat{H}_0^{\rm{e}}$) and perturbed ($\hat{H}'$) Hamiltonians
%
%
\begin{equation}
\hat{H}^{\rm{e}} =\hat{H}_0^{\rm{e}} + \hat{H}'.
\end{equation}
Here, $\hat{H}^{\mathrm{e}}_0$ is the atomic electronic Hamiltonian
%
%
\begin{equation}\label{ref:H0}
\hat{H}^{\mathrm{e}}_0=\sum_i\left[c \bm{\alpha} \cdot \bm{p}_i +(\beta-1)  c^2+V_{\mathrm{nuc}}\left(r_i\right)\right]+ \sum_{i \neq j} g(r_{i},r_{j}),
\end{equation}
where $c$ is the speed of light, $\bm{\alpha}$ and ${\beta}$ are Dirac matrices, $\bm{p}$ is the momentum operator, $V_{\mathrm{nuc}}\left(r_i\right)$ is the electron-nucleus interaction potential, and $g(r_{i},r_{j})$ is the two-electron interaction operator. When $V_{\mathrm{nuc}}$ is a spherical potential 
\footnote{The case where $V_{\mathrm{nuc}}$ is not spherical corresponds to molecules where the atomic states (atomic orbitals) with opposite parity can mix without external electric fields in the molecule-fixed frame. Instead, a (smaller) electric field has to be applied to mix the rotational or $\Lambda(\Omega)$-doublet states with opposite parity, in order to orient the molecule axis.
}, 
$\hat{H}_0$ is $\mathcal{P}$-even, and the zeroth-order wavefunction $|\Psi_m^{(0)}\rangle$ is $\mathcal{P}$-definite. $\hat{H}'$ consists of the contributions from the MQM and the lowest-order electric-dipole interaction of electrons with an externally applied static electric field,
%
%
\begin{equation}\label{eq:H_pert_MQM}
\hat{H}' =  \hat{H}_{\mathrm{MQM}}
- \sum_i \textbf{r}_i \cdot \mathbf{\mathcal{E}}^{\mathrm{ext} 
}.
\end{equation}
Note that the interaction term between the MQM and external fields does not exist. As a result, the first-order contribution for the energy shift $E^{(1)}_m$ is purely zero,
%
%
\begin{equation}
\begin{split}
E^{(1)}_m & =\left\langle\Psi_m^{(0)}\right|\hat{H}'\left|\Psi_m^{(0)}\right\rangle \\
& =\left\langle\Psi_m^{(0)}\right|\hat{H}_{\mathrm{MQM}} \left|\Psi_m^{(0)}\right\rangle
-\left\langle\Psi_m^{(0)}\right|  \sum_i \textbf{r}_i \cdot \mathbf{\mathcal{E}}^{\mathrm{ext}} \left|\Psi_m^{(0)}\right\rangle  
=0,
\end{split}
\end{equation}
where $|\Psi_m^{(0)}\rangle$ is the zeroth-order wavefunction of the unperturbed Hamiltonian $\hat{H}_0 $ given by,
\begin{equation}
\hat{H}_0\left|\Psi_m^{(0)}\right\rangle = E_m^{(0)} \left|\Psi_m^{(0)}\right\rangle.
\end{equation}
This is due to the fact that the MQM term $\hat{H}_{\mathrm{MQM}}$ and electric dipole $\textbf{r}_i$ are $\mathcal{P}$-odd, while $|\Psi_m^{(0)}\rangle$ is $\mathcal{P}$-definite. 

Next, we consider the second-order energy shift. The first-order wavefunction can be expressed by
%
%
\begin{equation}\label{eq:psi1}
\left|\Psi_m^{(1)}\right\rangle=\sum_{n \neq m} \frac{\left|\Psi_n^{(0)}\right\rangle\left\langle\Psi_n^{(0)}\left|\hat{H}'\right| \Psi_m^{(0)}\right\rangle}{E_m^{(0)}-E_n^{(0)}},
\end{equation}
When we retain only the terms that are proportional to $\mathcal{M}$ and ignoring terms proportional to $\mathcal{M}^2$, then the second-order energy shift in an atom is given by,
%
%
\begin{eqnarray}\label{eq:2th1}
E_{m}^{(2)} 
 =\sum_{n \neq m} \frac{\left\langle\Psi_m^{(0)}\left|\hat{H}_{\mathrm{MQM}}
\right| \Psi_n^{(0)}\right\rangle\left\langle\Psi_n^{(0)}|- \sum_i \mathbf{r}_i \cdot \mathbf{\mathcal{E}}^{\mathrm{ext}}| \Psi_m^{(0)}\right\rangle}{E_m^{(0)}-E_n^{(0)}} \nonumber\\
 +\sum_{n \neq m} \frac{\left\langle\Psi_n^{(0)}|- \sum_i 
 \mathbf{r}_i \cdot \mathbf{\mathcal{E}}^{\mathrm{ext}}| \Psi_m^{(0)}\right\rangle\left\langle\Psi_m^{(0)}\left|\hat{H}_{\mathrm{MQM}}\right| \Psi_n^{(0)}\right\rangle}{E_m^{(0)}-E_n^{(0)}}.
\end{eqnarray}
Here, we consider $\mathbf{\mathcal{E}}^{\mathrm{ext}} = \mathcal{E}_z \hat{z}$, so that
%
%
\begin{equation}\label{eq:parameter}
 E_m^{(2)}   
=\sum_{{n} \neq {m}}  \frac{\left\langle\Psi_{m}^{(0)}\left|
 \hat{H}_{\mathrm{MQM}}
\right| \Psi_{n}^{(0)}\right\rangle\left\langle\Psi_{n}^{(0)}|- \sum_i z_i | \Psi_{m}^{(0)}\right\rangle}{E_{m}^{(0)}-E_n^{(0)}} \, \mathcal{E}_z 
 +c.c.
\end{equation}
Now, the energy shift can be expressed as $E^{(2)}_m=-\left\langle D_{m} \right\rangle \mathcal{E}_z$, where
the atomic EDM $\left\langle D_{m} \right\rangle$ induced by the MQM can be expressed as follows:
%
%
\begin{eqnarray}\label{eq:atomic_MQM}
  \left\langle D_{m} \right\rangle =
\frac{1}{6} \frac{\mathcal{M}}{I(2 I-1)}\left[3 {M_I}^2-I(I+1)\right]  \nonumber\\
 \times  \sum_{{n} \neq {m}}  \frac{\left\langle\Psi_{m}^{(0)}\left|
\frac{3}{c} \sum_i \left(\frac{\mathbf{r}_i \times \boldsymbol{\alpha}_i}{r_i^5}\right)_zz_i
\right| \Psi_{n}^{(0)}\right\rangle\left\langle\Psi_{n}^{(0)}| \sum_j z_j | \Psi_{m}^{(0)}\right\rangle}{E_{m}^{(0)}-E_{n}^{(0)}}  +c.c.
\end{eqnarray}
%
%

The perturbative expression derived above is useful to clarify the physical background. However, our relativistic many-body calculations include the term of the electric dipole interaction with an external electric field in the unperturbed Hamiltonian, instead of the perturbation,  
%
\begin{equation}\label{eq:DCG_ext}
\hat{H}^\mathrm{e} = \hat{H}^\mathrm{e}_0 - \sum_i z_i \mathcal{E}_z.
\end{equation}
In this study, its eigenfunction $\left|\Psi^\mathrm{e}_{m}\right\rangle$ is computed based on the relativistic configuration interaction (CI) theory. The energy shift due to the MQM is
%
%
\begin{equation}\label{eq:val_MQM}
E^{\mathrm{e}}_{\mathrm{MQM},M_J} = -\frac{1}{6} \frac{\mathcal{M}}{I(2 I-1)}\left[3 {M_I}^2-I(I+1)\right] 
\left\langle\Psi^\mathrm{e}_{M_J}\left|
\frac{3}{c} \sum_i \left(\frac{\mathbf{r}_i \times \boldsymbol{\alpha}_i}{r_i^5}\right)_zz_i
\right| \Psi^\mathrm{e}_{M_J}\right\rangle.
\end{equation}
Here we replaced $m$ with $M_J$, which is the projection of the total electronic angular momentum $J$ along the quantization axis, to distinguish the ones in the latter section, where the hyperfine coupling is taken into account. With current technology, the maximum external electric field that can be applied in the MQM experiment is about 200 kV/cm $\sim 4 \times 10^{-4}$ a.u. \cite{Ready2021NIM_Ra}, which is well within the perturbative regime. 
Therefore the energy shift is proportional to the electric field $\mathcal{E}_z$ applied in the experiment to a very good approximation. This fact allows us to define the quantity 
%
\begin{equation}\label{eq:T}
T_{M_J} = \frac{1}{\mathcal{E}_z} \left\langle\Psi^\mathrm{e}_{M_J}\left|  \frac{3}{c} \sum_i \left(\frac{\mathbf{r}_i \times \boldsymbol{\alpha}_i}{r_i^5}\right)_zz_i\right| \Psi^\mathrm{e}_{M_J}\right\rangle  .
\end{equation}
which only depends on atomic properties and is independent of experimental parameters such as the applied electric field. $T_{M_J}$ is thus a measure of the intrinsic sensitivity of an atomic state to the MQM, analogous to the enhancement factor $R$ \cite{Das2017Handbook} (or $K$ \cite{Ginges2004PR,Khriplovich1997_strangeness}) used in atomic eEDM calculations.
Eq. \ref{eq:val_MQM} can be expressed in terms of $T_{M_J}$ as
%
%
\begin{equation}\label{eq:Jz_MQM_e_ene}
E^{\mathrm{e}}_{\mathrm{MQM},{M_J}} \approx -\frac{1}{6} \frac{\mathcal{M}}{I(2 I-1)}\left[3 {M_I}^2-I(I+1)\right]\, T_{M_J} \, \mathcal{E}_z.
\end{equation}

\subsubsection{The case of nonzero hyperfine coupling}\label{subsubsec:IJ_val}
Taking the hyperfine coupling into account, atomic states should be described by the total angular momentum $\hat{\textbf{F}}=\hat{\textbf{I}}+\hat{\textbf{J}}$. 
%
%
The total atomic Hamiltonian to obtain the MQM-energy shift in the $\left|FM_F\right>$ states is given by,
%
%
\begin{equation}\label{eq:full_Hamiltonian}
\hat{H}^\mathrm{A} = \hat{H}^\mathrm{A}_0 + \hat{H}^{J_z}_{\mathrm{MQM}}.
\end{equation}
Here, $\hat{H}^{J_z}_{\mathrm{MQM}}$ is the effective MQM operator that provides the matrix element described by Eq. \ref{eq:Jz_MQM_e_ene} in the $\left|JM_J\right>\left|IM_I\right>$ basis.
$\hat{H}^\mathrm{A}_0$ includes the hyperfine interactions, the Zeeman interaction of the electronic angular momentum and nuclear spin with a magnetic field as well as the direct current (DC) Stark shift 
 due to an external static electric field, and is given as,
%
\begin{eqnarray}\label{eq:atom_Hamiltonian}
\hat{H}^\mathrm{A}_0 
= A_{\mathrm{hfc}} \hat{\mathbf{I}} \cdot \hat{\mathbf{J}} 
+ B_{\mathrm{hfc}} \frac{3 (\hat{\mathbf{I}} \cdot \hat{\mathbf{J}})^2 + \frac{3}{2}(\hat{\mathbf{I}} \cdot \hat{\mathbf{J}})-\hat{\mathbf{I}}^2 \hat{\mathbf{J}}^2}{ 2 I(2 I-1) J(2 J-1)}  \nonumber \\
+ \left( g_J \mu_B \hat{J_z} -g_I \mu_N \hat{I_z}\right) \mathcal{B}_z  
- \frac{1}{2} \alpha^T_{J}(0) \frac{3 \hat{J}_z^2-\hat{\textbf{J}}^2}{J(2 J-1)} \mathcal{E}_z^2 .
\end{eqnarray}
%

Here $A_{\mathrm{hfc}}$ and $B_{\mathrm{hfc}}$ are the usual hyperfine structure coefficients; $g_J$ ($g_I$) is the $g$ factor of the total electron angular momentum (nuclear spin); $\mu_N$ and $\mu_B$ are the nuclear magneton and Bohr magneton, respectively.
The atomic parameters in Eq. (\ref{eq:atom_Hamiltonian}) for $^{173}$Yb in the $^3P_2$ state are summarized in Table \ref{tbl:parameters}. The corresponding eigenstates and eigenvalues are
%
\begin{equation}\label{eq:H_0}
\hat{H}^\mathrm{A}_0\left|\Psi^\mathrm{A}_{0,F,M_F}\right> = E^\mathrm{A}_{0,F,M_F}\left|\Psi^\mathrm{A}_{0,F,M_F}\right>.
\end{equation}

%
%
\begin{table}
\caption{Atomic parameters in Eq (\ref{eq:atom_Hamiltonian}) that are used in our study. 
}\label{tbl:parameters}
\begin{tabular}{lccccc}
\br
 & $A_{\mathrm{hfc}}$ (MHz) & $B_{\rm{hfc}}$ (MHz) & $g_I$ ($\mu_N$) & $g_J$ ($\mu_{{B}}$) & $\alpha^T_J(0)$ (a.u.) \tabularnewline
\mr
value & $-$742.7 & 1342 & $-$0.6776 & 1.5 & $-$76 \tabularnewline
reference & \cite{Wakui2003JPSP_YbHFCC} & \cite{Wakui2003JPSP_YbHFCC} & \cite{Porsev2004PRA_3P2_g} & Land\'e $g$-factor & \cite{Porsev1999PRA_alpha}\tabularnewline
\br
\end{tabular}

\end{table}

Since the hyperfine interaction is dominant in the $\hat{H}^\mathrm{A}_0$, $F$ and $M_F$ are good quantum numbers, and the energy shift from the effective MQM Hamiltonian can be obtained from the first-order perturbation theory, as follows:
%
\begin{equation}\label{eq:E_MQM}
\Delta E_\mathrm{MQM}({F,M_F})
= \left<\Psi^\mathrm{A}_{0,{F,M_F}}\left|\hat{H}^{J_z}_{\mathrm{MQM}} \right|\Psi^\mathrm{A}_{0,{F,M_F}}\right>.
\end{equation}
The MQM energy shifts for all hyperfine states are shown in Fig. \ref{fig:MQM_all}.
The presented values are obtained by carrying out the calculation described in the following subsections.  
The key feature is that the sign of the energy shift depends on the signs of $M_F$  and $\mathcal{E}_z$, reflecting the $\mathcal{T}$-odd and $\mathcal{P}$-odd nature of the MQM Hamiltonian.

\subsection{Computational details}\label{subsec:technical}
The electronic structure calculations were carried out using a development version (git hash f49c9b1) of the DIRAC program package \cite{DIRAC19,saue2020dirac}. 
All calculations were based on the Dirac-Coulomb-Gaunt Hamiltonian including the external electric field ($1.0\times10^{-5}$ a.u $\sim$ 50 kV/cm) in the form of the electric-dipole approximation to the quantization axis. 
The $\left<SS|SS\right>$-type integrals are explicitly computed. 
The Dyall basis set with valence triple-$\zeta$ quality (dyall v3z) \cite{Dyall_Gomes_2010} was employed in the uncontracted form, and the small components of the basis sets were generated based on the kinetic balance \cite{Stanton1984JCP}. Diffuse functions are added to $s$, $p$, $d$, and $f$ orbitals in an even-tempered manner.
We employed the KR-CI module \cite{Fleig2003JCP,Knecht2010JCP,Knecht_thesis} for the relativistic configuration-interaction calculations.
The Gaussian-type nuclear charge distribution \cite{visscher1997ADNDT} was employed for the electron-nuclear interaction potential. 

The atomic spinors, which are the reference state functions for the CI wavefunction, were generated using the average-of-configuration Hartree-Fock (AOC-HF) method \cite{Thyssen_thesis} at the electronic configuration [Xe]$4f^{14}6s^{1}6p^{1}$. We employed the generalized-active-space (GAS) method to obtain the CI wavefunctions, where the number of allowed holes can be specified for each shell. We employ three models shown in Tables \ref{tbl:GAS_space} and \ref{tbl:acronym}. Here, two holes are allowed in the $4d$, $5s$, $5p$, $6s$, $6p$ shells, while the truncation of the virtual spinors and the number of holes in the $4p$ shells are different between each model. To obtain the target electronic configuration [Xe]$4f^{14}6s^16p^1$ with certainty, we specified 1 as the maximum occupation number in $6s$ shell, which provides a hole forcibly in this shell. The remaining electron occupies the $6p$ shell. 

%
%
\begin{table}
\caption{Generalized active space models for the CI wave functions for the $^3P_2$ state. The parameters $k$ and $N$ determine the constraints of the occupation for each GAS shell. The space with $N$ virtual Kramers pairs consists of the canonical AOC-HF spinors. The cutoff energies corresponding to $N$ are provided in Table \ref{tbl:acronym}.}\label{tbl:GAS_space}
\begin{center}
\begin{tabular}{cccc}
\br
Orbital & \# of Kramers pairs & \multicolumn{2}{c}{accumulated \# of electrons}\tabularnewline
 &  & min & max\tabularnewline
 \mr
virtual & $N$ & 26 & 26\tabularnewline
6p & 3 & 24 & 26\tabularnewline
6s & 1 & 23 & 25\tabularnewline
4d5s5p & 9 & 22 & 24\tabularnewline
4p & 3 & 6$-k$ & 6\tabularnewline
fronzen & (22) &  & \tabularnewline
\br
\end{tabular}
\end{center}
\end{table}

%
%
\begin{table}
    \caption{Definition of the GASCI wavefunction models. The value of $L$ in the parentheses ($L$ a.u.) is the cutoff energy corresponding to the number of virtual spinors $N$. e1 (e2) means that even-tempered diffuse functions $1s1p1d1f$ ($2s2p2d2f$) are added to the dyall v3z basis sets. cv refers to the core-valence correlation.}\label{tbl:acronym}
\begin{center}
\begin{tabular}{ccccc}
\br
$k$ & $N_{\mathrm{v3z}}$ & $N_{\mathrm{v3z+e1}}$ & $N_{\mathrm{v3z+e2}}$ & correlation model acronym\tabularnewline
\mr
0 & 98 & 114 & 130 & CISD(10 a.u.)\tabularnewline
1 & 98 & 114 &  & cv-CISD(10 a.u.)\tabularnewline
0 & 114 & 130 &  & CISD(30 a.u.)\tabularnewline
\br
\end{tabular}
\end{center}
\end{table}

\subsection{Results of calculations}\label{subsec:results}
\subsubsection{Electronic structure calculations}\label{subsec:electronic_calculation}
The results of the computation of $T_{M_J}$ are summarized in Table \ref{tbl:T}. Although dyall.v3z includes the primitive functions for $6p$ orbitals, the effects of the diffuse functions are significant, especially for $M_J=2$ state. This is because $T_{M_J}$ is a response property to an external electric field, where diffuse functions are required to correctly describe the polarization of the wavefunction. The size of v3z+e1 basis sets is sufficient because the difference of $T_{M_J}$ between the v3z+e1 and v3z+e2 is negligible. The $M_J=2$ state is more sensitive to correlation effects than the $M_J=1$ state. The final values and uncertainties in Table \ref{tbl:T} are obtained as follows: 
\begin{eqnarray}\label{eq:final}
 \mathrm{Final} \; \mathrm{value} 
 &=& \mathrm{v3z+e1/cv\mathchar`-CISD(10 a.u.)}  \nonumber \\
 &+& \left[\mathrm{v3z+e1/CISD(30 a.u.)} - \mathrm{v3z+e1/CISD(10 a.u.)}\right] \nonumber \\
 &+& \left[\mathrm{v3z+e2/CISD(10 a.u.)} - \mathrm{v3z+e1/CISD(10 a.u.)}\right]
\end{eqnarray}
%
\begin{eqnarray}
 \mathrm{Uncertainty} \nonumber &=& \mathrm{abs[v3z+e1/cv\mathchar`-CISD(10 a.u.)} - \mathrm{v3z+e1/CISD(10 a.u.)]}\nonumber \\
 &+& \mathrm{abs[v3z+e1/CISD(30 a.u.)} - \mathrm{v3z+e1/CISD(10 a.u.)}] \nonumber \\
 &+& \mathrm{abs[v3z+e2/CISD(10 a.u.)} - \mathrm{v3z+e1/CISD(10 a.u.)}]
\end{eqnarray}

The final value for the $M_J=1$ subspace of $^3P_2$ is $T_{M_J} = -2.43(8)$ a.u., and the value in the $M_J=2$ subspace is $T_{M_J} = -2.85(21)$ a.u.

%
%
\begin{table}
    \caption{The values of $T_{M_J}$ (a.u.) of the Yb atom in $^3P_2$ state with various basis sets and different wavefunction models. The main text describes how the final values and uncertainty were estimated.}\label{tbl:T}
\begin{center}
\begin{tabular}{ccccccc}
\br
Basis & \multicolumn{2}{c}{v3z} & \multicolumn{2}{c}{v3z+e1} & \multicolumn{2}{c}{v3z+e2}\tabularnewline
$M_J$ & 1 & 2 & 1 & 2 & 1 & 2\tabularnewline
\mr
CISD(10 a.u.) & $-$1.98 & $-$0.83 & $-$2.51 & $-$2.65 & $-$2.51 & $-$2.64\tabularnewline
cv-CISD(10 a.u.) & $-$1.94 & $-$0.87 & $-$2.45 & $-$2.75 &  & \tabularnewline
CISD(30 a.u.) & $-$1.96 & $-$0.88 & $-$2.49 & $-$2.75 &  & \tabularnewline
Final &  &  & $-$2.43(8) & $-$2.85(21) &  & \tabularnewline
\br
\end{tabular}
\end{center}
\end{table}

Note that Yb has another open-shell state $4f^{13}6s^{2}5d$ ($J=2$) that has been recently observed \cite{Ishiyama2023PRL}. This state has a high sensitivity to various phenomena in new physics \cite{Shaniv2018PRL_LLI,Safronova2018PRL_alpha,Tang2023PRA}. According to our preliminary calculations, however, it is not sufficiently sensitive for an MQM search. The parameters $T_{M_J=1} = 0.13$ a.u. and $T_{M_J=2}= 0.16$ a.u. for this state, obtained with the relativistic CI method, are much smaller than for the $^3P_2$ state.

%
%
\subsubsection{MQM sensitivities}\label{subsubsec:MQM_sensitivities}
From the calculated values of $T_{M_J}$, we can evaluate the energy shift due to the MQM.
In Fig. \ref{fig:MQM_all}, we show the energy shift $\Delta E_{\mathrm{MQM}}({F,M_F})$ (Eq. \ref{eq:E_MQM}) for a magnetic quadrupole moment magnitude $\mathcal{M} = 1$ $\mu_N\mathrm{fm}$. Here, the eigenstate $\left|\Psi^\mathrm{A}_{0,{F,M_F}}\right>$ is expanded by the basis $\left|IM_I\right>\left|JM_J\right>$.
The magnetic moment $\mu_{F,M_F}$ \cite{Foot2004_atomphys}, a property that parametrizes the sensitivity of $\left|F,M_F\right>$ state to external magnetic field fluctuations, is given by 
\begin{equation}\label{eq:g_F}
\mu_{F,M_F}
= -g_F \mu_B M_F 
= -\left<\Psi^\mathrm{A}_{0,F,M_F}\right|
 g_J \mu_B \hat{J_z} - g_I \mu_N \hat{I_z}
 \left|\Psi^\mathrm{A}_{0,F,M_F}\right>.
\end{equation}
Here, $g_F$ is the g-factor. The computed values of $\mu_{F,M_F}$ are shown in Fig. \ref{fig:MQM_all}. 

We find that the maximum MQM sensitivity is not obtained in the state $M_F = F$, except for the $F=9/2$ states. 
A qualitative explanation of this effect is as follows: $E^{\mathrm{e}}_{\mathrm{MQM},M_J}$ given by Eq. \ref{eq:val_MQM} can be separated into nuclear ($I$ and $M_I$) and electronic ($M_J$) contributions. From Table \ref{tbl:T}, the electronic contribution is approximately common for $M_J=+1,+2$ states, and for  $M_J=-1,-2$ states. The nuclear part ($3 M_I^2-I(I+1)$) is $-8,-2,10$ for $|M_I|=1/2,3/2,5/2$, respectively. Therefore, the largest MQM shift appears in states where only $|M_I|=5/2$ contributes, such as $\left|F=9/2,M_F=\pm9/2\right>$. In the other states, the sensitivity depends on the Clebsch–Gordan (CG) coefficient $\left<IM_IJM_J|IJFM_F\right>$, and therefore the most sensitive nuclear spin sublevels $|I M_I=\pm5/2,JM_J\neq 0>$ does not always have a large contribution to a given $\left|F, M_F = \pm F\right>$ state. 

%
\section{Experimental scheme}\label{sec:experiment}
From an experimental perspective, it is desirable to perform spectroscopic measurements between pairs of hyperfine states that have large differential values of $\Delta E_{\mathrm{MQM}}({F,M_F})$ and small differential $\mu_{F,M_F}$.

Further, among the many possible combinations of available hyperfine state pairs, transitions between states with the the same $|M_F|$ values are useful, because of the insensitivity of their energy differences to the tensor light shift from the optical trapping field, as well as tensor shifts due to the applied static electric field.
In particular, a large difference $\Delta E_{\mathrm{MQM}}(F,M_F)$ and the smallest differential magnetic moment can be obtained using the transition between $\left|F=3/2, M_F=1/2\right> \rightarrow\left|F=3/2, M_F=-1/2\right>$ (see Fig. \ref{fig:MQM_all}). The reason for the large differential MQM shift between these states is that the CG coefficient $\left<IM_I=\pm 5/2,JM_J=\mp 2|IJFM_F=\pm 1/2\right>$ is largest when $F=3/2$ and smallest when $F=9/2$.

This pair of states is also preferred because they can be connected by a single-photon radio frequency (RF) transition. These states ($F=3/2$) can also be conveniently populated from the ground electronic $^1S_0$ ($F=5/2$) state in Yb using a hyperfine-induced $E1$ transition \cite{Boyd2007PRA_HFCC_E1}.



%
%
An important source of technical noise and fluctuations in an MQM measurement is the instability of the magnetic field applied to the atoms. In this respect, small sensitivity of the transition frequency to magnetic field fluctuations is advantageous. The energy of a hyperfine state is not simply proportional to the magnetic field, due to the competition between the hyperfine and Zeeman interactions. This fact gives rise to an interesting situation where the resonance frequency of particular transition can be made insensitive to magnetic fields by finding the condition ${\partial E^\mathrm{A}_{0,F,M_F}}/{\partial \mathcal{B}^{\mathrm{ext}}}={\partial E^\mathrm{A}_{0,F',M_F'}}/{\partial \mathcal{B}^{\mathrm{ext}}}$. 
%
\begin{figure*}[ht]
\includegraphics[width=0.9\textwidth]{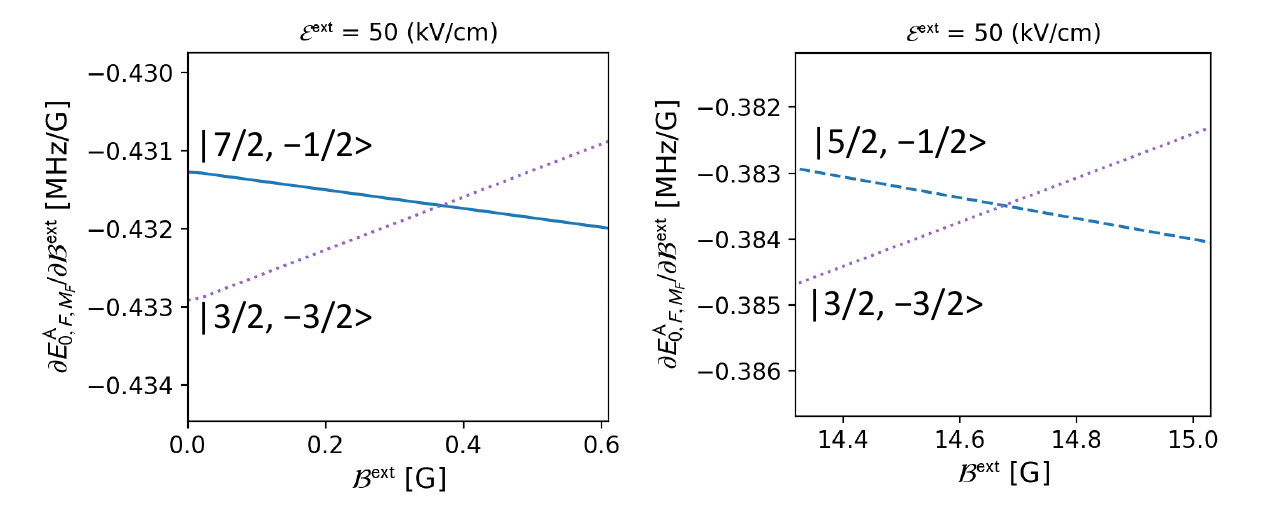}
\caption{\label{fig:dE_dB} Examples of the magic magnetic fields under $\mathcal{E}^{\mathrm{ext}}=50$ kV/cm. The left (right) one corresponds to the transition between $\left|7/2,-1/2\right>$ and $\left|3/2,-3/2\right>$ ($\left|5/2,-1/2\right>$ and $\left|3/2,-3/2\right>$) states at $B^\mathrm{ext}=0.37$ G ($B^\mathrm{ext}=14.7$ G). The other transitions are shown in Table \ref{tbl:magic}.}
\end{figure*}
We refer to such a condition as a ``magic magnetic field'' (MMF). Similar ideas have been explored in the context of eEDM search with nearly-degenerate states in polyatomic molecules \cite{Anderegg2023arXiv_zero_g} and in the more general context of molecular CP-violation searches with transitions between non-degenerate states, and in the latter case have shown their general existence among many molecular species \cite{Takahashi2023arXiv}.
 The possibility of MMFs are investigated in the experimentally convenient region $\mathcal{B} < 20$ G, and the MMF values are summarized in Table \ref{tbl:magic}, where we find several MMFs for $\mathcal{E}^\mathrm{ext}$ values of 50 kV/cm and 100 kV/cm. Two examples of MMFs at $\mathcal{E}^{\mathrm{ext}}=50$ kV/cm are shown in Fig. \ref{fig:dE_dB}, where the magnetic field sensitivity is plotted as a function of magnetic field. 
%
%
%
\begin{table}
    \caption{Summary of the magic magnetic fields $\mathcal{B}^{\mathrm{ext}}$ (G) that provide the same gradient of the energy with respect to the magnetic field  ${\partial E^\mathrm{A}_{0,F,M_F}}/{\partial \mathcal{B}^{\mathrm{ext}}}$ (MHz/G) for the two states, $\left|F,M_F\right>$ and $\left|F',M'_F\right>$. $\Delta E$ (GHz) is the energy difference between the two states. $\Delta E_{\mathrm{MQM}}$ (Hz/$\mu_N\mathrm{fm}$) is the expected differential MQM shift between the two states assuming $\mathcal{M}$ = 1 $\mu_N\mathrm{fm}$, obtained from Eq. \ref{eq:E_MQM}. Two different static electric fields $\mathcal{E}^{\mathrm{ext}}$ = 50, 100 (kV/cm) are considered. $\delta \mathcal{M}$ is the sensitivity to the MQM ($\times 10^{-8}\mu_N\mathrm{fm}$) for a shot-noise-limited measurement. Bold labels indicate magnetic-dipole-allowed single-photon transitions between states.}\label{tbl:magic}
\centering
\begin{tabular}{cccc ll cc}
\br
$\mathcal{E}^{\mathrm{ext}}$  & $\mathcal{B}^{\mathrm{ext}}$ & ${\partial E_{0,F,M_F}}/{\partial \mathcal{B}^{\mathrm{ext}}}$ & $\Delta E$  & $\left|F,M_F\right>$ & $\left|F',M'_F\right>$  & $\Delta E_{\mathrm{MQM}}$& $\delta \mathcal{M}$\tabularnewline
\mr
50 & 0.37 & $-$0.43 & 5.35 & $\left|7/2,-1/2\right>$ & $\left|3/2,-3/2\right>$  
& 81& 24.2\tabularnewline
& 5.03 & 0.45 & 7.80 & $\left|9/2,1/2\right>$ & $\left|3/2,3/2\right>$ 
&  97& 20.3 \tabularnewline
& 14.5 & $-$1.31 & 7.08 & $\left|7/2,-3/2\right>$ & $\left|1/2,1/2\right>$  
& 299 & 6.6 \tabularnewline
& \textbf{14.7} & $-$0.38 & 2.51 & $\left|5/2,-1/2\right>$ & $\left|3/2,-3/2\right>$ 
&  80& \textbf{24.5 }\tabularnewline
   \hline
100 & 5.3 & $-$0.46 & 7.82 & $\left|9/2,-1/2\right>$ & $\left|3/2,-3/2\right>$ 
&  195&  10.1\tabularnewline
& 10.7 & $-$0.44 & 5.33 & $\left|7/2,-1/2\right>$ & $\left|3/2,-3/2\right>$ 
&  166& 11.8 \tabularnewline
& 16.2 & $-$1.37 & 9.57 & $\left|9/2,-3/2\right>$ & $\left|1/2,1/2\right>$ 
&  536& 3.7 \tabularnewline
\br
\end{tabular}

\end{table}

In very small external magnetic fields, the quantization axis could potentially be disturbed by light shifts from the trapping light used in the tweezer array or optical lattice. For our expected experimental parameters, the Stark shift from the static electric field is significantly larger than the light shift, and thus the electric field direction, parallel to the magnetic field, robustly defines the quantization axis.

The proposed MQM measurement can be carried out using ultracold atoms in an optical lattice \cite{Schafer2020NP_QS} 
or an optical tweezer array \cite{Kaufman2021NP_tweezer}, 
which has several practical advantages for MQM measurements compared to experiments with beams or vapour cells. One advantage of the optical lattice and optical tweezer systems is the small volume occupied by the trapped-atom ensemble, as a result of which the inhomogeneities of magnetic and electric fields are usually very small. Trapping atoms in ultra-high vacuum conditions easily allows the application of electric fields up to 100 kV/cm.  Cold atoms in traps are also ideal platforms to apply advanced quantum metrology methods such as spin squeezing, enabling more precise measurements than the standard quantum limit. Spin squeezing methods for cold atoms, including Yb, have been demonstrated \cite{Takano2009PRL_squeezing,Takano2010PRL_nonclassical,Inoue2013PRL_quantum-noise,Bornet2023arXiv_dipolar_Rydberg,Penafiel2020Nature_entanglement,Eckner2023arXiv_squeezing_rydberg}. 
In addition, a Heisenberg-limited measurement could be possible by using the Greenberger-Horne-Zeilinger state \cite{Bollinger1996PRA_GHZ}, which has been successfully created for an optical tweezer array \cite{Omran2019science_GHZ}. 
This yields measurement uncertainty proportional to $1/N$, rather than $1/\sqrt{N}$ in the case of the standard quantum limit, where $N$ is the number of atoms.

Our group has demonstrated the preparation of more than $10^4$ atoms of $^{173}$Yb in the ground state $^1S_0$ at several tens of nano-kelvin, and successful loading of the atoms into a 3D optical lattice, resulting in the formation of a Mott-insulating state where typically one atom is occupied in one site \cite{Taie2012NP_SU6,Taie2022NP_SUN}. 
The efficient excitation of Yb atoms to the $^3P_2$ state in a 3D optical lattice has been demonstrated \cite{Tomita2019PRA_2body_loss}. 
When we use the transitions between the different $|M_F|$ states or different $F$ states for the MQM measurements, as in Table \ref{tbl:magic}, tensor light shifts are important sources of systematic and statistical uncertainties.
However, we can work with a magic wavelength for trapping laser light where the tensor light shift vanishes.
Our preliminary calculations based on reported values \cite{Tang2018JPB_E1,NIST_ASD,Hara2014JPSJ} suggest the existence of such a wavelength around 900 nm for the $^3P_2$ state.

Recent experiments from our group have also demonstrated the preparation of Yb atoms in the ground state $^1S_0$ in each of 50 sites of an optical tweezer array, followed by excitation to the $^3P_2$ state \cite{Okuno2022JPSP}. We anticipate the preparation of $10^3$ atoms of $^{173}$Yb in the $^3P_2$ state is possible. This is justified by the estimation of necessary tweezer light power of about 10 mW per site and the fact that the laser power of more than 10 W  at 532 nm is commercially available.
Since the radiative lifetime of the $^3P_2$ state is 10 s \cite{Porsev2004PRA_3P2_g}, we expect the coherence time between hyperfine states can be made comparable, by employing magnetic field stabilization and dynamical decoupling pulses to stabilize the field at the micro-gauss level. As discussed above, working at a magic magnetic field value can further relax the constraints on magnetic field stability required to achieve 10 s coherence time.

We can use the energy level shift coefficients computed in this paper to estimate the experimental sensitivity to the MQM. For an experiment limited by quantum projection noise from the atomic ensemble (``shot-noise limit''), the statistical uncertainty for the transition frequency is given by $\delta f = 1/\sqrt{N\tau T}$, where $N$ is the number of atoms, $\tau$ is the coherence time, and $T$ is the total measurement time. Using conservative experimental parameters, $N=10^3$, $\tau=1$ s and, $T=2.6\times 10^6$ s (thirty days), the frequency precision is $\delta f \approx 2.0\times10^{-5}$ Hz. The corresponding sensitivity to the MQM is:
%
\begin{equation}\label{eq:upper_limit}
\delta \mathcal{M} = \frac{\delta f}{\Delta E_{\mathrm{MQM}}}.
\end{equation}
The MQM sensitivity values are listed in Table \ref{tbl:magic}. 
The best sensitivity in this table is 
$\delta \mathcal{M} = 3.7\times 10^{-8}$ $\mu_N\mathrm{fm}$, under the condition $\mathcal{B}^{\mathrm{ext}}=16.2$ G and $\mathcal{E}^{\mathrm{ext}}=100 $ kV/cm.
In the case of the transition $\left|F=3/2, M_F=1/2\right> \rightarrow\left|F=3/2, M_F=-1/2\right>$ in the condition of Fig. \ref{fig:MQM_all}, $\delta \mathcal{M} = 5.2\times 10^{-8}$ $\mu_N\mathrm{fm}$ is expected. These estimates indicate that an experiment using $^{173}$Yb in the $^3P_2$ state has sufficient precision to improve upon the current experimental bound $\mathcal{M} <1.6\times10^{-6}$ $\mu_N\mathrm{fm}$ obtained using Cs \cite{Murthy1989PRL_Cs}. 

\section{Summary}\label{sec:conclusion}
We have described a competitive method to search for nuclear MQMs using $^{173}$Yb atoms in the $^3P_2$ state trapped in an optical lattice or optical tweezer array, and computed relevant atomic properties such as the MQM enhancement, hyperfine magnetic moments and magic magnetic field conditions.
Our estimated sensitivity $\delta \mathcal{M} \leq 3.7\times 10^{-8}$ $\mu_N\mathrm{fm}$ has the potential to improve upon the current experimental limit by more than one order of magnitude. 
This proposal focuses on $^{173}$Yb atoms in the metastable $^3P_2$ state, but the methods and ideas described here are generally applicable to new physics searches with other atoms and molecules.


\ack
This work was supported by the Grant-in-Aid for Scientific Research of JSPS (No. JP17H06138, No. JP18H05405, No. JP18H05228, No. JP22K20356), JST CREST (Nos. JPMJCR1673 and JPMJCR23I3), and MEXT Quantum Leap Flagship Program (MEXT Q-LEAP) Grant No. JPMXS0118069021, and JST Moonshot R\&D Grant No. JPMJMS2269.
A.S. acknowledges financial support from the JSPS KAKENHI Grant no. 21K14643. In this work, A.S. used the computer resource offered under the category of General Projects by the  Research Institute for Information Technology, Kyushu University, and the FUJITSU Supercomputer PRIMEHPC FX1000 and FUJITSU Server PRIMERGY GX2570 (Wisteria/BDEC-01) at the Information Technology Center, The University of Tokyo. A.V.\ gratefully acknowledges support from a JSPS Invitational Fellowship for Research in Japan (No. S22097) hosted by Kyoto University. Y. T. was supported by the Masason Foundation.

\begin{flushleft}
$^{*}$ sunagaayaki@gmail.com
\end{flushleft}

\bibliographystyle{iopart-num}
\bibliography{sunaga}

\end{document}